\documentclass[floatfix,twocolumn,prl,amsmath]{revtex4-2}
\usepackage{graphicx}
\usepackage{bm}
\usepackage{amssymb}
\usepackage{dcolumn}
\usepackage{subfigure}
\usepackage{multirow}
\usepackage{mathrsfs}
\usepackage{threeparttable}
\DeclareMathAlphabet{\mathscrbf}{OMS}{mdugm}{b}{n}
\usepackage{color}

\begin{document}
\newcommand{\intqspa}{\int\!\!\frac{\rmd^d q}{(2\pi)^d}}
\newcommand{\intqspathr}{\int\!\!\frac{\rmd^3 q}{(2\pi)^3}}
\newcommand{\intqspatwo}{\int\!\!\frac{\rmd^2 q}{(2\pi)^2}}
\newcommand{\intkspatwo}{\int\!\!\frac{\rmd^2 k}{(2\pi)^2}}
\newcommand{\intkspa}{\int\!\!\frac{\rmd^d k}{(2\pi)^d}}
\newcommand{\intkspapri}{\int\!\!\frac{\rmd^d k'}{(2\pi)^d}}
\newcommand{\vn}[1]{{\boldsymbol{#1}}}
\newcommand{\vht}[1]{{\boldsymbol{#1}}}
\newcommand{\matn}[1]{{\bf{#1}}}
\newcommand{\matnht}[1]{{\boldsymbol{#1}}}
\newcommand{\bege}{\begin{equation}}
\newcommand{\gretke}{G_{\vn{k} }^{\rm R}(\mathcal{E})}
\newcommand{\gret}{G^{\rm R}}
\newcommand{\gadv}{G^{\rm A}}
\newcommand{\gmat}{G^{\rm M}}
\newcommand{\gles}{G^{<}}
\newcommand{\ghat}{\hat{G}}
\newcommand{\sigmahat}{\hat{\Sigma}}
\newcommand{\glesone}{G^{<,{\rm I}}}
\newcommand{\glestwo}{G^{<,{\rm II}}}
\newcommand{\gspec}{G^{\rm S}}
\newcommand{\glesthree}{G^{<,{\rm III}}}
\newcommand{\magdir}{\hat{\vn{n}}}
\newcommand{\sigmaret}{\Sigma^{\rm R}}
\newcommand{\sigmales}{\Sigma^{<}}
\newcommand{\sigmalesone}{\Sigma^{<,{\rm I}}}
\newcommand{\sigmalestwo}{\Sigma^{<,{\rm II}}}
\newcommand{\sigmalesthree}{\Sigma^{<,{\rm III}}}
\newcommand{\sigmaadv}{\Sigma^{A}}
\newcommand{\ee}{\end{equation}}
\newcommand{\bal}{\begin{aligned}}
\newcommand{\defbar}{\overline}
\newcommand{\SM}{\scriptstyle}
\newcommand{\rmd}{{\rm d}}
\newcommand{\rme}{{\rm e}}
\newcommand{\eal}{\end{aligned}}
\newcommand{\crea}[1]{{c_{#1}^{\dagger}}}
\newcommand{\annihi}[1]{{c_{#1}^{\phantom{\dagger}}}}
\newcommand{\udot}{\overset{.}{u}}
\newcommand{\exponential}[1]{{\exp(#1)}}
\newcommand{\phandot}[1]{\overset{\phantom{.}}{#1}}
\newcommand{\phandag}{\phantom{\dagger}}
\newcommand{\Trace}{\text{Tr}}
\setcounter{secnumdepth}{2}
\title{Construction of Wannier functions from the spectral moments of correlated electron systems}
\author{Frank Freimuth$^{1,2}$}
\email[Corresp.~author:~]{f.freimuth@fz-juelich.de}
\author{Stefan Bl\"ugel$^{1}$}
\author{Yuriy Mokrousov$^{1,2}$}
\affiliation{$^1$Peter Gr\"unberg Institut and Institute for Advanced Simulation,
Forschungszentrum J\"ulich and JARA, 52425 J\"ulich, Germany}
\affiliation{$^2$ Institute of Physics, Johannes Gutenberg University Mainz, 55099 Mainz, Germany
}
\begin{abstract}
  When the first four spectral moments are considered, spectral features missing
  in standard Kohn-Sham (KS) density-functional theory (DFT), such as
  upper and lower Hubbard bands, as well as spectral satellite peaks,
  can be described, and the bandwidths can be corrected.
  Therefore, we have devised a \textit{moment-functional based spectral
  density functional theory} (MFbSDFT) recently.
  However, many computational tools in theoretical solid state physics,
  such as the construction
  of maximally localized Wannier functions (MLWFs),
  have been developed for KS-DFT and require
  modifications if they are supposed to be used in MFbSDFT. Here, 
  we show how generalized  Wannier functions may be constructed from the
  first four spectral moment matrices. We call these functions \textit{maximally
  localized spectral moment Wannier functions} (MLSMWFs).
  We demonstrate how MLSMWFs may be used to compute the anomalous Hall
  effect (AHE) in fcc Ni by Wannier interpolation.
  More generally, MLSMWFs may be computed from the first $2P$ moments ($P=1,2,3,\dots$).
  Using more than 4 moments opens the perspective of reproducing all spectral features accurately in MFbSDFT.
\end{abstract}
\maketitle
\section{Introduction}

Maximally localized Wannier functions (MLWFs)
have become a widely-applied tool in
computational solid state physics.
Recent reviews~\cite{rmp_wannier90,wannier90communitycode}
give a comprehensive overview of their
applications.
Using Wannier interpolation~\cite{wannierinterpolation}
one may significantly reduce 
the computing time requirements
of calculations of response functions
such as the
anomalous Hall effect (AHE)~\cite{PhysRevB.74.195118},
thermoelectric
coefficients~\cite{boltzwann},
and
the spin-orbit torque (SOT)~\cite{ibcsoit,invsot}.

The Wannier interpolation of the AHE
does not suffer from band truncation errors,
because it directly interpolates the Berry curvature.
In order to interpolate the SOT
and the Dzyaloshinskii-Moriya interaction
without
band truncation errors one may use
higher-dimensional Wannier functions~\cite{PhysRevB.91.184413}
in order to interpolate the mixed Berry
curvature~\cite{PhysRevB.101.014428,hdwfs_dmi}.
Response functions that cannot be
expressed in terms of these geometric properties of
the electronic bands cannot be treated within
the conventional Wannier interpolation method
without any band truncation error. However,
for such response functions Wannier function perturbation
theory has been developed recently~\cite{PhysRevX.11.041053}.

All these MLWF-based interpolation methods have been
devised essentially in the context of KS-DFT.
While impressive progress has been achieved in developing
exchange-correlation functionals for KS-DFT that describe
ground state properties of solids with high precision~\cite{dur36734},
experimental spectra are often not reproduced well by
standard KS-DFT~\cite{beyond_dft}.
In Ref.~\cite{spectral}
and in 
Ref.~\cite{momentis}
we have explained how the
spectral function may be
constructed from the first
four spectral moment matrices.
These moment matrices may
be obtained either 
by computing several correlation
functions self-consistently~\cite{spectral},
or by evaluating suitable moment
functionals~\cite{momentis}.
We expect that the moment potentials
required for the latter approach, which we
call MFbSDFT, are universal functionals of the
spin density.

Indeed, we have shown that 
parameter-free moment functionals can be found
that improve the spectra of Na and SrVO$_3$
significantly~\cite{momdis} in comparison to standard KS-DFT with LDA.
In order to formulate these parameter-free moment
functionals we have used an existing model
of the second moment of the uniform electron
gas (UEG)~\cite{PhysRevB.69.045113}
as well as models of the momentum distribution function
$n_{\vn{k}}$ of the UEG~\cite{PhysRevB.66.235116,PhysRevB.50.1391,PhysRevB.56.9970}.
However, formulating universal moment functionals that
are generally applicable is still a long way. Notably,
the correct inclusion of spin-polarization and the extension
by gradient corrections are important necessary developments
left for future work.

Nevertheless, even while the universal moment functionals are not
yet available, one may generally improve spectra by optimizing the
parameters in suitable parameterized moment functionals. 
Adjusting in this way e.g.\ the bandwidths and gaps in order to reproduce
experimental data will increase the accuracy of response function calculations, in
particular those of optical responses, such as laser-induced
currents~\cite{PhysRevResearch.4.043046}
and torques~\cite{PhysRevB.103.054403},
which are expected to be generally sensitive to gaps, bandwidths,
and band positions, because they are not Fermi-surface effects like the
AHE~\cite{PhysRevLett.93.206602,PhysRevB.76.195109,PhysRevB.89.117101}
for example.

We have demonstrated that MFbSDFT can be used to improve the description 
of the electronic structure of fcc Ni significantly in comparison to LDA,
because it yields bandwidth, exchange splitting,
and satellite peak positions in good agreement with
the experimental spectrum~\cite{momentis,momdis}.
The satellite peak in Ni roughly 6~eV below the
Fermi energy is a correlation
effect~\cite{PhysRevLett.43.1431,nickel_PhysRevB.40.5015,nickel_Borgiel1990},
which is missing in the
KS-spectrum. Since the method of spectral  moments
captures the splitting of bands into lower and upper Hubbard bands,
it cannot be mapped onto a non-interacting effective KS Hamiltonian in general
without changing the spectrum.
The question therefore poses itself of how to obtain generalized Wannier
functions from the spectral moment matrices.

The method of spectral moments yields state energies, state wavefunctions,
and corresponding spectral weights,
in contrast to KS-DFT, where the spectral
weights are by definition unity.
Moreover, the state wavefunctions corresponding to different energies
are not guaranteed  to be orthogonal.
While
many-body generalizations of Wannier
functions have been considered
before~\cite{PhysRevB.79.045109},
these generalizations are not optimized to construct
localized Wannier functions from the first four spectral moment matrices.
For example, Ref.~\cite{PhysRevB.79.045109}
does not take the spectral weights of the quasiparticles into account and does not
consider the possibility that the quasiparticle wavefunctions do not necessarily
form a set of mutually orthogonal functions. However, the standard
method of constructing MLWFs assumes the Bloch functions of different bands
to be orthogonal~\cite{PhysRevB.65.035109,PhysRevB.56.12847}.
Here, we will show that spectral weights and non-orthogonality
of quasiparticle wavefunctions can be taken into account by generalizing the
MLWFs concept for the method of spectral moments.

We demonstrate the construction of MLSMWFs
for fcc Ni and use them to compute the
AHE by Wannier interpolation. 
We choose fcc Ni because it is known that standard KS-DFT
overestimates the bandwidth and the exchange splitting in
this material~\cite{nickel_PhysRevB.40.5015}
and it predicts the AHE to be significantly larger
than the experimental
value~\cite{PhysRevLett.107.106601,PhysRevB.87.060406,PhysRevB.84.144427}.
Moreover, even the sign of the
magnetic anisotropy energy (MAE) is not predicted correctly by standard
KS-DFT, i.e., LDA does not predict the correct easy axis~\cite{PhysRevLett.87.216405}.
Previously, we have demonstrated that MFbSDFT 
can be used to reproduce the experimental values of the
exchange splitting, the bandwidth, and the position of the satellite
peaks~\cite{momentis,momdis}.
Here, we demonstrate that also the AHE is predicted to be close to the
experimental value if MFbSDFT is used.

The rest of this paper is structured as follows.
In Sec.~\ref{sec_theory}
we explain how we construct MLSMWFs from the first
four spectral moment matrices.
Practical issues, such as the use of the {\tt wannier90}
code~\cite{wannier90communitycode}
for the generation of MLSMWFs are described in
Sec.~\ref{sec_overlaps} and in Sec.~\ref{sec_initial_proj}.
MLSMWFs with spin-orbit interaction (SOI) are
discussed in Sec.~\ref{sec_mlsmwfs_soi}.
In 
Sec.~\ref{sec_implementation_flapw}
we explain that
\textit{ab-initio} programs which can compute MLWFs can
be extended easily to compute additionally MLSMWFs.
In Sec.~\ref{sec_interpolate_mlsmwfs_ahe}
we describe Wannier interpolation based on MLSMWFs using the example
of the AHE. The interpolation of additional matrix elements such as
spin and torque operators is discussed in Sec.~\ref{eq_interpolate_spin_torque}.
In Sec.~\ref{sec_more}
we explain how the method of Sec.~\ref{sec_theory}
may be generalized to include the first $2P$ moments ($P=3,4,\dots$).
In Sec.~\ref{sec_results} we present applications
of our method to fcc Ni.
This paper ends with a summary in Sec.~\ref{sec_summary}.

\section{Theory}
\label{sec_theory}
Before discussing the construction of MLSMWFs from the first four spectral
moment matrices in Sec.~\ref{sec_mlwfs_from_moments}
we first revisit the
generation of MLWFs in Sec.~\ref{sec_revisit_mlwfs}
as well as the calculation of the spectral
function in Sec.~\ref{sec_revisit_specfun}.
This will help us to explain the
necessary modifications of the MLWFs formalism in Sec.~\ref{sec_mlwfs_from_moments},
when MLSMWFs are constructed from the spectral moment matrices.
\subsection{MLWFs and Wannier interpolation}
\label{sec_revisit_mlwfs}
The Bloch functions $|\psi_{\vn{k}m}\rangle$
are eigenfunctions of the KS-Hamiltonian $H$
with eigenenergies $\mathcal{E}_{\vn{k}m}$:
\bege
\label{eq_ks_hamilton}
H|\psi_{\vn{k}m}\rangle
=
\mathcal{E}_{\vn{k}m}
|\psi_{\vn{k}m}\rangle,
\ee
where $\vn{k}$ is the $\vn{k}$-point and $m$ is the band index.
The MLWFs $|W_{\vn{R}n}\rangle$  are constructed from these
Bloch functions 
by the transformation~\cite{PhysRevB.65.035109}
\bege\label{eq_define_mlwf}
|W_{\vn{R}n}\rangle=\frac{1}{\mathcal{N}}\sum_{\vn{k}}\sum_{m=1}^{N_{\rm B}} U_{mn}^{(\vn{k})} e^{-i\vn{k}\cdot \vn{R}} |\psi_{\vn{k}m}\rangle,
\ee
where $\mathcal{N}$ is the number of $\vn{k}$ points. 
The matrix $\vn{U}^{(\vn{k})}$ is a
rectangular matrix when the number
of bands $N_{\rm B}$ is larger than the number of MLWFs $N_{\rm W}$,
otherwise it is a square matrix, which may occur for example
when MLWFs are constructed from isolated groups of bands~\cite{PhysRevB.56.12847}.
The matrix $\vn{U}^{(\vn{k})}$ is determined by the condition
that the MLWFs minimize the spatial spread
\bege\label{eq_spread_mlwfs}
\Omega=\sum_{n=1}^{N_{\rm W}}
\left[
\langle
W_{\vn{R}n}
|
\vn{r}^2
|
W_{\vn{R}n}
\rangle
-\left(
\langle
W_{\vn{R}n}
|
\vn{r}
|
W_{\vn{R}n}
\rangle\right)^2
\right].
\ee

Due to the localization of the MLWFs in real-space
the matrix elements of the Hamiltonian $H$ decay rapidly
when the distance between the MLWFs increases:
\bege
\label{eq_loc_ks}
\lim_{\vn{R}\rightarrow \infty}
\langle
W_{\vn{0}m}
|
H
|
W_{\vn{R}n}
\rangle
\rightarrow 0.
\ee
This localization property is important for Wannier
interpolation, because it implies that
in order to interpolate
the electronic
structure at any desired $\vn{k}$ point
it is sufficient to provide
the matrix elements $\langle
W_{\vn{0}m}
|
H
|
W_{\vn{R}n}
\rangle$ for a finite and small set of $\vn{R}$ vectors.
The reason for this is that the electronic structure at
any desired $\vn{k}$ point may be computed by performing
a Fourier transformation of $\langle
W_{\vn{0}m}
|
H
|
W_{\vn{R}n}
\rangle$ and that the computational time for this is small if the
set of $\vn{R}$ vectors is small.
Explicitly, the matrix elements of the
Hamiltonian may be written as
\bege\label{eq_hamil_mat_wbas_ks}
\begin{aligned}
&\langle
W_{\vn{0}m}
|
H
|
W_{\vn{R}n}
\rangle=\\
&=\frac{1}{\mathcal{N}}
\sum_{\vn{k}}
\sum_{m'=1}^{N_{\rm B}}
e^{-i\vn{k}\cdot\vn{R}}
\mathcal{E}_{\vn{k}m'}
\left[U^{(\vn{k})}_{m'm}\right]^{*}
U^{(\vn{k})}_{m'n}.
\end{aligned}
\ee

In order to describe the AHE in magnetically collinear ferromagnets,
the spin-orbit interaction (SOI) has to be taken into account.
Wannier interpolation is very efficient in computing the
AHE~\cite{PhysRevB.74.195118}.
In the presence of SOI, the Bloch functions
are spinors
\bege
\langle
\vn{r}
|
\psi_{\vn{k}m}
\rangle=
\psi_{\vn{k}m}(\vn{r})=
\begin{pmatrix}
  \chi_{\vn{k}m\uparrow}(\vn{r})\\
   \chi_{\vn{k}m\downarrow}(\vn{r})
\end{pmatrix},
\ee
and the MLWFs
are spinors as well:
\bege
\label{eq_spinor_mlwfs}
\langle
\vn{r}
|
W_{\vn{R}n}
\rangle=
\begin{pmatrix}
  W_{\vn{R}n\uparrow}(\vn{r})\\
  W_{\vn{R}n\downarrow}(\vn{r})
\end{pmatrix}.
\ee
For the interpolation of the AHE in bcc Fe~\cite{PhysRevB.74.195118}
one
first constructs MLWFs using a coarse $\vn{k}$ mesh,
e.g.\ an $8\times 8\times 8$ mesh.
Next, one computes the matrix elements of the
Hamiltonian according to Eq.~\eqref{eq_hamil_mat_wbas_ks}.
Finally, one may Fourier transform these matrix elements
for all  $\vn{k}$ points in the fine interpolation mesh,
e.g.\ an $800\times 800 \times 800$ mesh. Using this interpolated
Hamiltonian, one may compute the AHE numerically
efficiently~\cite{PhysRevB.74.195118}. 

\subsection{Construction of the spectral function from the first four spectral moment matrices}
\label{sec_revisit_specfun}
In Ref.~\cite{momentis} we describe an algorithm
to construct the spectral function from the
first four spectral moments.
In the following we assume that the spectral
moments are so expressed in a basis set of $N_{\rm S}$ orthonormalized
functions $\phi_{\vn{k}n}(\vn{r})$
that they are given by
$N_{\rm S}\times N_{\rm S}$ matrices. Due to the orthonormalization
of the basis functions, the zeroth moment $\vn{M}_{\vn{k}}^{(0)}$
is simply the unit matrix.
When the spectral function is determined approximately
from the first four spectral moments ($\vn{M}_{\vn{k}}^{(0)}$, $\vn{M}_{\vn{k}}^{(1)}$, $\vn{M}_{\vn{k}}^{(2)}$, and $\vn{M}_{\vn{k}}^{(3)}$),
the poles of the
single-particle spectral function are
given by the eigenenergies of the hermitean
$2N_{\rm S}\times 2N_{\rm S}$ matrix~\cite{momentis}
\bege\label{eq_2n2n_mat}
\vn{\mathcal{H}}_{\vn{k}}=\left(
\begin{array}{cc}
\vn{M}_{\vn{k}}^{(1)}
&\vn{B}_{1\vn{k}} \\
\vn{B}_{1\vn{k}}^{\dagger} &\vn{D}_{1\vn{k}}
\end{array}
\right),
\ee
where $\vn{M}_{\vn{k}}^{(1)}$
is the first moment,
$\vn{B}_{1\vn{k}}=\vn{\mathcal{U}}_{\vn{k}}\sqrt{\vn{\mathcal{D}}_{\vn{k}}}$,
$\vn{B}_{2\vn{k}}=[\vn{M}_{\vn{k}}^{(3)}-\vn{M}_{\vn{k}}^{(2)}\vn{M}_{\vn{k}}^{(1)}][\vn{B}_{1\vn{k}}^{\dagger}]^{-1}$,
and
$\vn{D}_{1\vn{k}}=\vn{B}_{1\vn{k}}^{-1}[\vn{B}_{2\vn{k}}-\vn{M}_{\vn{k}}^{(1)}\vn{B}_{1\vn{k}}]$.
Here, $\vn{\mathcal{D}}_{\vn{k}}$ is a diagonal matrix,
and $\vn{\mathcal{U}}_{\vn{k}}$ is a unitary matrix
so that $\vn{\mathcal{U}}_{\vn{k}}\vn{\mathcal{D}}_{\vn{k}}\vn{\mathcal{U}}_{\vn{k}}^{\dagger}=\vn{M}_{\vn{k}}^{(2)}-\vn{M}_{\vn{k}}^{(1)}\vn{M}_{\vn{k}}^{(1)}$.

The eigenvectors of $\vn{\mathcal{H}}_{\vn{k}}$, Eq.~\eqref{eq_2n2n_mat},
may be written as
\bege\label{eq_state_vectors}
\vn{\Psi}_{\vn{k}n}=\begin{pmatrix}
\vn{\psi}_{\vn{k}n\rightarrow}\\
\vn{\psi}_{\vn{k}n\leftarrow}
\end{pmatrix},
\ee
where $\vn{\psi}_{\vn{k}n\rightarrow}$ and $\vn{\psi}_{\vn{k}n\leftarrow}$
are both column vectors with $N_{\rm S}$ components,
while $\vn{\Psi}_{\vn{k}n}$ is a column vector with $2N_{\rm S}$ entries.
We denote the eigenvalues of $\vn{\mathcal{H}}_{\vn{k}}$
by $\mathscr{E}_{\vn{k}n}$, i.e.,
\bege\label{eq_bigspecmat_eigprobl}
\vn{\mathcal{H}}_{\vn{k}}
\vn{\Psi}_{\vn{k}n}
=
\mathscr{E}_{\vn{k}n}\vn{\Psi}_{\vn{k}n}.
\ee
Within MFbSDFT, the charge density is computed only from the
upper part $\vn{\psi}_{\vn{k}n\rightarrow}$ of the
state vector $\vn{\Psi}_{\vn{k}n}$ (Eq.~\eqref{eq_state_vectors}), while
the lower part plays the role of an auxiliary component.
Note that while the eigenfunctions $|\psi_{\vn{k}m}\rangle$
of the KS-Hamiltonian (Eq.~\eqref{eq_ks_hamilton})
are orthonormal, i.e.,
\bege
\langle
\psi_{\vn{k}m}
|
\psi_{\vn{k}'n}
\rangle
=\delta_{nm}
\delta_{\vn{k}\vn{k}'},
\ee
the upper parts  $\vn{\psi}_{\vn{k}n\rightarrow}$ of the state
vectors $\vn{\Psi}_{\vn{k}n}$ (Eq.~\eqref{eq_state_vectors}) are not
even orthogonal, i.e.,
\bege
\left[\vn{\psi}_{\vn{k}m\rightarrow}\right]^{\dagger}
\vn{\psi}_{\vn{k}n\rightarrow}\not\propto \delta_{nm},
\ee
while the complete state vectors are
orthonormal:
\bege
\left[\vn{\Psi}_{\vn{k}m}\right]^{\dagger}
\vn{\Psi}_{\vn{k}n}= \delta_{nm}.
\ee

We may obtain the spectral weight of the
state $\vn{\Psi}_{\vn{k}n}$ from
\bege\label{eq_specwei}
a_{\vn{k}n}=
\left[\vn{\psi}_{\vn{k}n\rightarrow}\right]^{\dagger}
\vn{\psi}_{\vn{k}n\rightarrow}=
\sum_{m'=1}^{N_{\rm S}} [\vn{\psi}_{\vn{k}nm'\rightarrow}]^{*}
\vn{\psi}_{\vn{k}nm'\rightarrow}.
\ee
Here, $\psi_{\vn{k}nm'\rightarrow}$ is the $m'$-th entry in the
column vector $\vn{\psi}_{\vn{k}n\rightarrow}$.
These spectral weights are useful to quantify the
relative importance of a given state with energy $\mathscr{E}_{\vn{k}n}$.
For example, it may occur that the spectral function has a pole
at $\mathscr{E}_{\vn{k}n}$ with a spectral weight $a_{\vn{k}n}\ll 1$.
Due to the small spectral weight this pole might not be observable in the
experimental spectrum. Therefore, both the poles $\mathscr{E}_{\vn{k}n}$
and the spectral weights $a_{\vn{k}n}$ are generally necessary to discuss the spectrum.

In order to construct MLSMWFs from the state vectors $\vn{\Psi}_{\vn{k}n}$,
we need their real-space representation $\Psi_{\vn{k}n}(\vn{r})$.
Clearly, $\vn{\psi}_{\vn{k}n\rightarrow}$ is given by
\bege\label{eq_real_space_upper}
\psi_{\vn{k}n\rightarrow}(\vn{r})=
\sum_{m=1}^{N_{\rm S}}\phi_{\vn{k}m}(\vn{r})\psi_{\vn{k}nm\rightarrow},
\ee
in real-space,
where
$\phi_{\vn{k}m}(\vn{r})$ is the $m$-th function in the orthonormal set
used to express the spectral moments at $\vn{k}$.

In MFbSDFT the functions $\psi_{\vn{k}n\rightarrow}(\vn{r})$
replace the KS wavefunctions from standard KS-DFT~\cite{momentis}:
The charge density and the DOS may be obtained
from $\psi_{\vn{k}n\rightarrow}(\vn{r})$, while the auxiliary
vector $\vn{\psi}_{\vn{k}n\leftarrow}$ is only needed
to solve Eq.~\eqref{eq_bigspecmat_eigprobl}, and
may often be discarded afterwards.
However, as will become clear in Sec.~\ref{sec_mlwfs_from_moments},
we need the lower part $\psi_{\vn{k}n\leftarrow}(\vn{r})$
for the construction of MLSMWFs.
The matrices $\vn{B}_{1\vn{k}}$ and $\vn{B}_{2\vn{k}}$
describe linear maps from the space of eigenfunctions
of $\vn{M}_{\vn{k}}^{(2)}-\vn{M}_{\vn{k}}^{(1)}\vn{M}_{\vn{k}}^{(1)}$
to the space of orthonormal basis functions $\phi_{\vn{k}n}(\vn{r})$.
Consequently, the matrix $\vn{D}_{1\vn{k}}$
describes a linear map from the space of eigenfunctions
of $\vn{M}_{\vn{k}}^{(2)}-\vn{M}_{\vn{k}}^{(1)}\vn{M}_{\vn{k}}^{(1)}$
to itself. Therefore, the components of $\vn{\psi}_{\vn{k}n\leftarrow}$
refer to the space of eigenfunctions
of $\vn{M}_{\vn{k}}^{(2)}-\vn{M}_{\vn{k}}^{(1)}\vn{M}_{\vn{k}}^{(1)}$ and
an additional unitary transformation to the space of orthonormal
basis functions $\phi_{\vn{k}n}(\vn{r})$ is necessary to obtain the
real-space representation of $\vn{\psi}_{\vn{k}n\leftarrow}$:
\bege\label{eq_lower_in_real_space}
\psi_{\vn{k}n\leftarrow}(\vn{r})=
\sum_{m,m'=1}^{N_{\rm S}}\mathcal{U}_{\vn{k}mm'}\phi_{\vn{k}m}(\vn{r})\psi_{\vn{k}nm'\leftarrow}.
\ee

Another approach leading to Eq.~\eqref{eq_lower_in_real_space}
considers the unitary transformation
\bege
\mathscrbf{U}_{\vn{k}}=
\begin{pmatrix}
  \vn{1}&\vn{0}\\
  \vn{0}&\vn{\mathcal{U}}_{\vn{k}}
\end{pmatrix},  
\ee
where $\vn{1}$ is the $N_{\rm S}\times N_{\rm S}$ unit matrix,
while $\vn{0}$ is the $N_{\rm S}\times N_{\rm S}$ zero matrix.
When this transformation is applied to $\vn{\mathcal{H}}_{\vn{k}}$
it does not change its eigenvalues $\mathscr{E}_{\vn{k}n}$ nor the
upper part $\vn{\psi}_{\vn{k}n\rightarrow}$ of the eigenvectors.
Only the lower part $\vn{\psi}_{\vn{k}n\leftarrow}$ of the eigenvectors
is changed
so that Eq.~\eqref{eq_bigspecmat_eigprobl}
turns into
\bege\label{eq_tilde_h_eigenvalue_eq}
\bar{\vn{\mathcal{H}}}_{\vn{k}}
\begin{pmatrix}
\vn{\psi}_{\vn{k}n\rightarrow}\\
\vn{\mathcal{U}}_{\vn{k}}\vn{\psi}_{\vn{k}n\leftarrow}
\end{pmatrix}
=
\mathscr{E}_{\vn{k}n}
\begin{pmatrix}
\vn{\psi}_{\vn{k}n\rightarrow}\\
\vn{\mathcal{U}}_{\vn{k}}\vn{\psi}_{\vn{k}n\leftarrow}
\end{pmatrix},
\ee
where
\bege
\bar{\vn{\mathcal{H}}}_{\vn{k}}=
\mathscrbf{U}_{\vn{k}}
\vn{\mathcal{H}}_{\vn{k}}
\mathscrbf{U}_{\vn{k}}^{\dagger}
\ee
is the transformed Hamiltonian.
$\bar{\vn{\mathcal{H}}}_{\vn{k}}$
describes a
map $\mathscr{V}_{\vn{k}}\times \mathscr{V}_{\vn{k}}\rightarrow \mathscr{V}_{\vn{k}}\times \mathscr{V}_{\vn{k}}$,
where we denote the space of orthogonal
basis functions $\phi_{\vn{k}n}(\vn{r})$
by $\mathscr{V}_{\vn{k}}$.
Since the lower components of the eigenvectors of $\bar{\vn{\mathcal{H}}}_{\vn{k}}$
are given by $\vn{\mathcal{U}}_{\vn{k}}\vn{\psi}_{\vn{k}n\leftarrow}$
according to Eq.~\eqref{eq_tilde_h_eigenvalue_eq},
it is clear that the real-space representation of $\vn{\psi}_{\vn{k}n\leftarrow}$
is given by Eq.~\eqref{eq_lower_in_real_space}.

\subsection{Choice of the moment functionals}

The spectral moment matrices $\vn{M}_{\vn{k}}^{(I)}$ ($I=1,2,3,\dots$) may
be obtained either
by computing several correlation
functions self-consistently~\cite{spectral},
or by evaluating suitable moment
functionals~\cite{momentis,momdis}.
In the latter approach, which we call MFbSDFT,
the $I$-th moment is decomposed into
the $I$-th power of the first moment plus the
additional contribution $\vn{M}_{\vn{k}}^{(I+)}$~\cite{momentis,momdis}:
\bege
\vn{M}_{\vn{k}}^{(I)}=
\left[
  \vn{M}_{\vn{k}}^{(1)}
  \right]^{I}
+ \vn{M}_{\vn{k}}^{(I+)}.
\ee
The first moment, $\vn{M}_{\vn{k}}^{(1)}$, may be obtained
easily within the standard KS framework:
It is simply given by the KS Hamiltonian, if instead of the
full exchange-correlation potential only the first-order exchange
is used.
The additional contributions $\vn{M}_{\vn{k}}^{(I+)}$ may
be computed from suitable
potentials $\mathcal{V}^{(I+)}(\vn{r})$~\cite{momentis,momdis}:
\bege\label{eq_def_mi+}
M_{\vn{k} nm}^{(I+)}=\int d^3 r \mathcal{V}^{(I+)}(\vn{r}) \phi_{\vn{k}n}^{*}(\vn{r})\phi_{\vn{k}m}(\vn{r}).
\ee

We expect that the $\mathcal{V}^{(I+)}(\vn{r})$ depend only on the
electron density $n(\vn{r})$, i.e., there are universal functionals
of $n(\vn{r})$, from which $\mathcal{V}^{(I+)}(\vn{r})$ may be
obtained.
This expectation is corroborated by our finding~\cite{momdis}
that parameter-free expressions for the moment potentials can be
found that improve the spectra of Na and of SrVO$_3$ significantly
in comparison to standard KS-DFT with LDA.
However, general and accurate expressions for $\mathcal{V}^{(I+)}(\vn{r})$
are currently not yet available. Therefore,
we proposed several parameterizations of $\mathcal{V}^{(I+)}(\vn{r})$,
which can be used to reproduce spectral features such as satellite peaks
and to correct the band width e.g.\ in Ni~\cite{momentis,momdis}. 

Defining the  dimensionless density parameter
\bege\label{eq_dimlessdenparam}
r_s(\vn{r})=
\frac{1}{a_{\rm B}}
\left(
\frac{3}{4\pi n(\vn{r})}
\right)^{\frac{1}{3}},
\ee
where $a_{\rm B}$ is Bohr's radius,
we may express $\mathcal{V}^{(I+)}(\vn{r})$
through~\cite{momentis}
\bege\label{eq_expand_vi+}
\mathcal{V}^{(I+)}(\vn{r})=\frac{c^{(I+)}}{[r_s(\vn{r})]^{I}}+\dots 
\ee
in the low-density limit, i.e., when $r_s(\vn{r})$ is large.
Alternatively, one may use~\cite{momentis,momdis}
\bege\label{eq_expand_vi+_vc}
\mathcal{V}^{(I+)}(\vn{r})=d^{(I+)}
 [V_{c}(r_s)]^{I}+\dots, 
\ee
where
\bege
V_{c}=\frac{d(\epsilon_c n)}{dn}
\ee
is the correlation potential, and
$\epsilon_{c}$ is the correlation energy. 

In these parameterized expressions of $\mathcal{V}^{(I+)}$ one may
treat e.g.\ $d^{(2+)}$ and $d^{(3+)}$ as independent parameters and optimize
both in order to match the experimental spectra as well as possible.
Alternatively, one may compute $\mathcal{V}^{(3+)}$ for a given
$\mathcal{V}^{(2+)}$ by enforcing the constraint of the momentum 
distribution function of the UEG~\cite{momdis}. 

\subsection{Construction of MLSMWFs from the first four spectral moment matrices}
\label{sec_mlwfs_from_moments}
In order to compute MLSMWFs from the first four spectral moment matrices,
we need to use the states Eq.~\eqref{eq_state_vectors}
instead of the usual Bloch functions. 
An obvious generalization of Eq.~\eqref{eq_define_mlwf} based on
these state vectors is
\bege\label{eq_def_mlsmwfs}
\begin{pmatrix}
\langle \vn{r}|\mathcal{W}_{\vn{R}n\rightarrow}\rangle\\
\langle \vn{r}|\mathcal{W}_{\vn{R}n\leftarrow}\rangle
\end{pmatrix}
=\frac{1}{\mathcal{N}}
\sum_{\vn{k}}\sum_{m=1}^{2N_{\rm S}} U_{mn}^{(\vn{k})} e^{-i\vn{k}\cdot \vn{R}}
\begin{pmatrix}
  \psi_{\vn{k}m\rightarrow}(\vn{r})\\
  \psi_{\vn{k}m\leftarrow}(\vn{r})
\end{pmatrix}  
\ee
where
$\psi_{\vn{k}m\rightarrow}(\vn{r})$ and $\psi_{\vn{k}m\leftarrow}(\vn{r})$
are given
in Eq.~\eqref{eq_real_space_upper}
and Eq.~\eqref{eq_lower_in_real_space},
respectively,
and the $2N_{\rm S}\times N_{\rm W}$ matrix
$\vn{U}^{(\vn{k})}$
is so chosen that
the spread
\bege\label{eq_spead_mlsmwfs}
\begin{aligned}
\Omega=&\sum_{n=1}^{N_{\rm W}}
\left[
\langle
\mathcal{W}_{\vn{R}n\rightarrow}
|
\vn{r}^2
|
\mathcal{W}_{\vn{R}n\rightarrow}
\rangle
-\left(
\langle
\mathcal{W}_{\vn{R}n\rightarrow}
|
\vn{r}
|
\mathcal{W}_{\vn{R}n\rightarrow}
\rangle\right)^2
\right]\\
+&\sum_{n=1}^{N_{\rm W}}
\left[
\langle
\mathcal{W}_{\vn{R}n\leftarrow}
|
\vn{r}^2
|
\mathcal{W}_{\vn{R}n\leftarrow}
\rangle
-\left(
\langle
\mathcal{W}_{\vn{R}n\leftarrow}
|
\vn{r}
|
\mathcal{W}_{\vn{R}n\leftarrow}
\rangle\right)^2
\right]\\
\end{aligned}
\ee
is
minimized.

As a consequence of the spatial localization,
the matrix
elements $\langle \mathcal{W}_{\vn{0}m}
|
\mathcal{H}
|
\mathcal{W}_{\vn{R}n}
\rangle$
decay rapidly in real-space similar
to Eq.~\eqref{eq_loc_ks}:
\bege
\label{eq_loc_genwan}
\lim_{\vn{R}\rightarrow \infty}
\langle
\mathcal{W}_{\vn{0}m}
|
\mathcal{H}
|
\mathcal{W}_{\vn{R}n}
\rangle
\rightarrow 0.
\ee
Explicitly, these matrix elements
are given by
\bege\label{eq_hamilmat_genwan}
\begin{aligned}
  &\mathcal{H}_{\vn{R}mn}=
  \langle\mathcal{W}_{\vn{0}m}
|
\mathcal{H}
|
\mathcal{W}_{\vn{R}n}
\rangle
=\\
&=\frac{1}{\mathcal{N}}
\sum_{\vn{k}}
\sum_{m'=1}^{2N_{\rm S}}
e^{-i\vn{k}\cdot\vn{R}}
\mathscr{E}_{\vn{k}m'}
\left[U^{(\vn{k})}_{m'm}\right]^{*}
U^{(\vn{k})}_{m'n}.
\end{aligned}
\ee
The derivation of Eq.~\eqref{eq_hamilmat_genwan}
shows clearly that both $\psi_{\vn{k}n\rightarrow}(\vn{r})$
and $\psi_{\vn{k}n\leftarrow}(\vn{r})$
need to be taken into account in the construction of MLSMWFs:
Only when both components,  $\vn{\psi}_{\vn{k}n\rightarrow}$
and $\vn{\psi}_{\vn{k}n\leftarrow}$, are considered, $\vn{\Psi}_{\vn{k}n}$
is an eigenvector of $\vn{\mathcal{H}}_{\vn{k}}$.
Moreover, it is clear that 
both components,
$\langle \vn{r}|\mathcal{W}_{\vn{R}n\rightarrow}\rangle$
and
$\langle \vn{r}|\mathcal{W}_{\vn{R}n\leftarrow}\rangle$,
have to be localized together to minimize
Eq.~\eqref{eq_spead_mlsmwfs}, because otherwise
Eq.~\eqref{eq_loc_genwan}
is not valid and the Fourier transformation below 
in Eq.~\eqref{eq_ft_generalizedwan} cannot be performed numerically efficiently.

In order to obtain the interpolated band structure,
we first carry out the Fourier transformation
\bege\label{eq_ft_generalizedwan}
\tilde{\vn{\mathcal{H}}}_{\vn{k}}=\sum_{\vn R}
e^{i \vn{k}\cdot \vn{R}}
\vn{\mathcal{H}}_{\vn{R}},
\ee
where $\mathcal{H}_{\vn{R}}$
is the matrix with the components $\mathcal{H}_{\vn{R}mn}$ defined
in Eq.~\eqref{eq_hamilmat_genwan}.
Next, we diagonalize $\tilde{\vn{\mathcal{H}}}_{\vn{k}}$:
\bege\label{eq_diag_genwan_interpo}
[\vn{\mathcal{X}}_{\vn{k}}]^{\dagger}
\tilde{\vn{\mathcal{H}}}_{\vn{k}}\vn{\mathcal{X}}_{\vn{k}}=\tilde{\mathscrbf{E}}_{\vn{k}}.
\ee
Here, $\vn{\mathcal{X}}_{\vn{k}}$ 
is a unitary matrix and $\tilde{\mathscrbf{E}}_{\vn{k}}$
is a diagonal matrix holding the interpolated energies:
\bege
\tilde{\mathscr{E}}_{\vn{k}nm}=\tilde{\mathscr{E}}_{\vn{k}n}\delta_{nm}.
\ee

Often, we would like to interpolate not only the band energies
but also the spectral weights $a_{\vn{k}n}$, Eq.~\eqref{eq_specwei}.
For this purpose, we first need to compute the matrix elements
\bege\begin{aligned}
s_{\vn{k}nm}&=
(\vn{\psi}^{\dagger}_{\vn{k}n\rightarrow},\vn{\psi}^{\dagger}_{\vn{k}n\leftarrow})
\begin{pmatrix}
  \vn{1}&\vn{0}\\
  \vn{0} &\vn{0}
\end{pmatrix}
\begin{pmatrix}
  \vn{\psi}_{\vn{k}m\rightarrow}\\
  \vn{\psi}_{\vn{k}m\leftarrow}
\end{pmatrix}\\
&=\vn{\psi}^{\dagger}_{\vn{k}n\rightarrow}\vn{\psi}_{\vn{k}m\rightarrow}
\end{aligned}
\ee
for all $\vn{k}$ points in the coarse $\vn{k}$ mesh that are used
in the construction of the MLSMWFs.
Here, $\vn{0}$ is the $N_{\rm S}\times N_{\rm S}$ zero matrix
and $\vn{1}$ is the $N_{\rm S}\times N_{\rm S}$ unit  matrix.
Next, these matrix elements need to be expressed in the MLSMWF basis:
\bege\label{eq_srnm}
s_{\vn{R}nm}=\frac{1}{\mathcal{N}}\sum_{\vn{k}}\sum_{n',m'=1}^{2 N_{\rm S}}s_{\vn{k}n'm'}
\left[\mathcal{U}^{(\vn{k})}_{n'n}\right]^{*}
\mathcal{U}^{(\vn{k})}_{m'm}e^{-i\vn{k}\cdot\vn{R}}.
\ee
After carrying out these preparations before the actual
Wannier interpolation step, one may interpolate
$s_{\vn{k}nm}$ to a given $\vn{k}$ point in the fine
interpolation mesh by
performing the Fourier transformation
\bege\label{eq_sknm}
\tilde{s}_{\vn{k}nm}=\sum_{\vn{R}}s_{\vn{R}nm}e^{i\vn{k}\cdot \vn{R}}
\ee
in the course of the Wannier interpolation.
Finally, $\tilde{s}_{\vn{k}nm}$ needs to be transformed into
the eigenbasis in order to obtain the interpolated spectral weights:
\bege\label{eq_specwei_inter_final}
\tilde{a}_{\vn{k}n}=\sum_{n'm'}\tilde{s}_{\vn{k}n'm'}\mathcal{X}_{\vn{k}m'n}
\left[
\mathcal{X}_{\vn{k}n'n}
\right]^{*}.
\ee

While the applications shown below in
Sec.~\ref{sec_results} use the MFbSDFT
approach of Ref.~\cite{momentis} in order to
obtain the spectral moments,
the theory for the construction of the
MLSMWFs from the spectral moment matrices
that we present here
can also be used when the spectral moments
are obtained by computing several correlation
functions self-consistently as in Ref.~\cite{spectral}.

\subsection{Wavefunction overlaps}
\label{sec_overlaps}
The {\tt wannier90} code~\cite{wannier90communitycode}
computes the
spread Eq.~\eqref{eq_spread_mlwfs}
from
the overlaps between the lattice periodic parts
$u_{\vn{k} m}(\vn{r}) = e^{-i\vn{k}\cdot\vn{r}}\psi_{\vn{k} m}(\vn{r})$
of the Bloch functions
at the nearest-neighbor  $k$-points $\vn{k}$ and $\vn{k}+\vn{b}$.
Therefore, the matrix elements
\bege
M_{mn}^{(\vn{k},\vn{b})} =\langle u_{\vn{k} m}|u_{\vn{k}+\vn{b}, n}\rangle
\ee
need to be provided to {\tt wannier90}
in order to determine the MLWFs through the matrix $U^{(\vn{k})}_{mn}$
in Eq.~\eqref{eq_define_mlwf},
which minimizes the spread Eq.~\eqref{eq_spread_mlwfs}.

In order to find the matrix $U^{(\vn{k})}_{mn}$
that defines the MLSMWFs in
Eq.~\eqref{eq_def_mlsmwfs}
one may use the {\tt wannier90} code~\cite{wannier90communitycode}
as well. In this case one needs to provide the matrix elements
\bege\label{eq_mmnk_mlsmwfs}
\begin{aligned}
  M_{mn}^{(\vn{k},\vn{b})} =\langle u_{\vn{k} m\rightarrow}|u_{\vn{k}+\vn{b}, n\rightarrow}\rangle
  +\langle u_{\vn{k} m\leftarrow}|u_{\vn{k}+\vn{b}, n\leftarrow}\rangle
\end{aligned}
\ee
to {\tt wannier90}, which ensures that all contributions to the
spread in Eq.~\eqref{eq_spead_mlsmwfs}
are taken into account.

\subsection{Initial projections}
\label{sec_initial_proj}
In order to obtain a good starting point for the
iterative minimization of the spreads, Eq.~\eqref{eq_spread_mlwfs}
(for MLWFs)
and Eq.~\eqref{eq_spead_mlsmwfs} (for MLSMWFs), one may
define first guesses $|g_{n}\rangle$ for these
Wannier functions~\cite{PhysRevB.65.035109,PhysRevB.56.12847}.
In the case of MLWFs the matrix elements
\bege\label{eq_initial_proj}
A_{mn}^{(\vn{k})}=\langle\psi_{\vn{k}m}|g_{n}\rangle
\ee
may be computed and provided to the {\tt wannier90}
code~\cite{wannier90communitycode} for this purpose.
In order to provide the first guesses in the case of
MLSMWFs, one may generalize Eq.~\eqref{eq_initial_proj}
as follows:
\bege\label{eq_initial_proj_mlsmwfs}
A_{mn}^{(\vn{k})}=\langle\psi_{\vn{k}m\rightarrow}|g_{n\rightarrow}\rangle+
\langle\psi_{\vn{k}m\leftarrow}|g_{n\leftarrow}\rangle.
\ee

When one computes MLWFs of bulk transition metals such as bcc Fe,
fcc Ni, fcc Pt, and fcc Pd, one typically constructs
9 MLWFs per spin in order to obtain Wannier functions that describe the
valence bands and the first few conduction bands.
In this case suitable initial projections are one $s$,
three $p$, and five $d$ states, which are 9 states in total.
Alternatively, one may use 6 $sp^{3}d^{2}$ hybrid states
plus $d_{xy}$, $d_{yz}$, and $d_{zx}$.
From Sec.~\ref{sec_mlwfs_from_moments} it follows that
the number of MLSMWFs is typically chosen twice as large as the number of MLWFs would
be chosen in the same material.
If we assume that for half of the MLSMWFs the $\rightarrow$-component is
larger than the $\leftarrow$-component, while for the remaining other
half of the MLSMWFs
the $\leftarrow$-component is more dominant than the $\rightarrow$-component,
an obvious choice for the initial projections is to use states
that are purely $\leftarrow$ or purely $\rightarrow$.

\subsection{Construction of MLSMWFs in systems with SOI}
\label{sec_mlsmwfs_soi}
In magnetically collinear systems without SOI, one typically
constructs MLWFs separately for spin-up and spin-down, i.e.,
one constructs two sets of MLWFs.
In the presence of SOI
this is not possible, because the Hamiltonian couples the spin-up
and spin-down bands. Consequently, only a single set of MLWFs
is constructed.
For example, in ferromagnetic fcc Ni one computes 9 spin-up
MLWFs and 9 spin-down MLWFs when SOI is not taken into account,
while one constructs 18 spinor-MLWFs (see Eq.~\eqref{eq_spinor_mlwfs})
when SOI is considered.

Analogously, only a single set of
MLSMWFs is constructed in systems with SOI.
In this case every MLSMWF has four components:
\bege
\langle \vn{r}|\mathcal{W}_{\vn{R}n}\rangle
=
\begin{pmatrix}  
  \langle \vn{r}|\mathcal{W}_{\vn{R}n\rightarrow\uparrow}\rangle\\
\langle \vn{r}|\mathcal{W}_{\vn{R}n\rightarrow\downarrow}\rangle\\  
\langle \vn{r}|\mathcal{W}_{\vn{R}n\leftarrow\uparrow}\rangle\\
\langle \vn{r}|\mathcal{W}_{\vn{R}n\leftarrow\downarrow}\rangle
\end{pmatrix}.
\ee
Similarly, the eigenvectors $\vn{\Psi}_{\vn{k}n}$ in
Eq.~\eqref{eq_bigspecmat_eigprobl}
have four components:
\bege
\vn{\Psi}_{\vn{k}n}
=
\begin{pmatrix}
  \vn{\psi}_{\vn{k}n\rightarrow}\\
  \vn{\psi}_{\vn{k}n\leftarrow}\\
\end{pmatrix}
=
\begin{pmatrix}
  \vn{\psi}_{\vn{k}n\rightarrow\uparrow}\\
  \vn{\psi}_{\vn{k}n\rightarrow\downarrow}\\
  \vn{\psi}_{\vn{k}n\leftarrow\uparrow}\\
  \vn{\psi}_{\vn{k}n\leftarrow\downarrow}
\end{pmatrix}.  
\ee

Consequently, the matrix elements
\bege\label{eq_mmnk_mlsmwfs_soi}
M_{mn}^{(\vn{k},\vn{b})} =\sum_{p=\rightarrow,\leftarrow}\sum_{\sigma=\uparrow,\leftarrow}
\langle u_{\vn{k} m p \sigma}|u_{\vn{k}+\vn{b}, n p \sigma}\rangle
\ee
and
\bege\label{eq_initial_proj_mlsmwfs_soi}
A_{mn}^{(\vn{k})}=
\sum_{p=\rightarrow,\leftarrow}\sum_{\sigma=\uparrow,\leftarrow}
\langle\psi_{\vn{k}m p \sigma}|g_{n p \sigma}\rangle
\ee
need to be provided to the {\tt wannier90} code in this
case in order to determine the MLSMWFs.

\subsection{Implementation within the FLAPW method}
\label{sec_implementation_flapw}
In Ref.~\cite{mlwfs_flapw}
we describe in detail how the matrix
elements $M_{mn}^{(\vn{k},\vn{b})}$
and $A_{mn}^{(\vn{k})}$ required by {\tt wannier90} for the
calculation of the MLWFs
may be implemented within the full-potential
linearized augmented plane-wave method (FLAPW).
For the construction of the
MLSMWFs
we need to compute these matrix elements
according to the
prescriptions of Eq.~\eqref{eq_mmnk_mlsmwfs}
and Eq.~\eqref{eq_initial_proj_mlsmwfs} (when SOI is
included in the calculations Eq.~\eqref{eq_mmnk_mlsmwfs_soi}
and Eq.~\eqref{eq_initial_proj_mlsmwfs_soi} should be used instead).
It is straightforward to extend the implementation described in
Ref.~\cite{mlwfs_flapw} by adding the additional loop over the MFbSDFT
indices $\rightarrow$ and $\leftarrow$.

\subsection{Wannier interpolation of response functions}
\label{sec_interpolate_mlsmwfs_ahe}
In Ref.~\cite{spectral}
we have described how the AHE conductivity may be computed
within the method of spectral moments using correlation functions
such as $\langle
    [[c^{\dagger}_{\vn{k}\alpha}c_{\vn{k}\beta},H]_{-},c^{\dagger}_{\vn{k}\gamma}c_{\vn{k}\delta}]_{-}
\rangle$ (see e.g.\ Eq.~(34), Eq.~(C1), and Eq.~(C2) in Ref.~\cite{spectral}).
However, we have also reported in Ref.~\cite{spectral} that
in the case of the Hubbard-Rashba model
the AHE is well-approximated by
\bege\label{eq_ahe_conduct}
\begin{aligned}
&\sigma_{xy}=\frac{e^2\hbar}{ V \mathcal{N}}
\sum_{\vn{k}}
  \sum_{n,n'=1}^{N_{\rm W}}
[f_{\vn{k}n}-f_{\vn{k}n'}]\times\\
&\times
\frac{{\rm Im}
\left[  
\langle \psi_{\vn{k}n\rightarrow} |  v_{x } |\psi_{\vn{k}n'\rightarrow}  \rangle
\langle \psi_{\vn{k}n'\rightarrow}  |  v_{y }| \psi_{\vn{k}n\rightarrow}   \rangle
\right]}
{(\mathscr{E}_{\vn{k}n'}-\mathscr{E}_{\vn{k}n})^2+0^{+}},
\end{aligned}
\ee
which does not require us to compute
correlation functions such as $\langle
    [[c^{\dagger}_{\vn{k}\alpha}c_{\vn{k}\beta},H]_{-},c^{\dagger}_{\vn{k}\gamma}c_{\vn{k}\delta}]_{-}
\rangle$.
Here, we assume that Eq.~\eqref{eq_ahe_conduct}
can also be used to compute the AHE of realistic materials approximately
within the spectral moment approach.
We leave if for future work to test
this approximation by computing the AHE also from the
correlation functions $\langle
    [[c^{\dagger}_{\vn{k}\alpha}c_{\vn{k}\beta},H]_{-},c^{\dagger}_{\vn{k}\gamma}c_{\vn{k}\delta}]_{-}
    \rangle$, and focus on the evaluation of
    Eq.~\eqref{eq_ahe_conduct}
in order to provide an example of Wannier interpolation with MLSMWFs.

We may obtain 
$\langle \psi_{\vn{k}n\rightarrow} |  v_{x } |\psi_{\vn{k}n'\rightarrow}  \rangle$
from
Wannier interpolation by
computing first the matrix elements
\bege
h_{\vn{k}nm}=\sum_{n',m'=1}^{N_{\rm S}} M^{(1)}_{n'm'}
\psi_{\vn{k}mm'\rightarrow}
\left[
  \psi_{\vn{k}nn'\rightarrow}
\right]^{*}
\ee
of the first moment for all $\vn{k}$ points
in the coarse $\vn{k}$ mesh
that are used
in the construction of the MLSMWFs.
Subsequently,
we compute the
corresponding matrix elements in the MLSMWFs basis:
\bege
h_{\vn{R}nm}=
\frac{1}{\mathcal{N}}\sum_{\vn{k}}
\sum_{n',m'=1}^{2 N_{\rm S}}h_{\vn{k}n'm'}
\left[\mathcal{U}^{(\vn{k})}_{n'n}\right]^{*}
\mathcal{U}^{(\vn{k})}_{m'm}e^{-i\vn{k}\cdot\vn{R}}.
\ee
These are preparatory steps that are carried out before the
actual Wannier interpolation.
In order to interpolate $h_{\vn{k}nm}$
to a given $\vn{k}$ point in the fine interpolation mesh
we first carry out the Fourier-transformation
\bege
\tilde{h}_{\vn{k}nm}=\sum_{\vn{R}}h_{\vn{R}nm}e^{i\vn{k}\cdot \vn{R}}.
\ee
The
velocity operator matrix elements are obtained from the $\vn{k}$ derivative:
\bege
\tilde{\vn{v}}_{\vn{k}nm}=
\frac{1}{\hbar}\sum_{\vn{R}}i\vn{R}e^{i\vn{k}\cdot \vn{R}}h_{\vn{R}nm}.
\ee
Finally, we need to transform these matrix elements into the
eigenbasis, which we obtain from Eq.~\eqref{eq_diag_genwan_interpo}:
\bege\label{eq_velo_inter_final}
\langle \psi_{\vn{k}n\rightarrow} |  \vn{v} |\psi_{\vn{k}m\rightarrow}  \rangle=
\sum_{n'm'}\tilde{\vn{v}}_{\vn{k}n'm'}\mathcal{X}_{\vn{k}m'm}
\left[
\mathcal{X}_{\vn{k}n'n}
\right]^{*},
\ee
where $\mathcal{X}_{\vn{k}m'm}$ are the elements of the unitary matrix
defined in
Eq.~\eqref{eq_diag_genwan_interpo}.
Now, the matrix elements Eq.~\eqref{eq_velo_inter_final}
may be used together with the eigenvalues $\mathscr{E}_{\vn{k}n}$
and the Fermi factors $f_{\vn{k}n}=f(\mathscr{E}_{\vn{k}n})$ (where $f$ is the
Fermi function) to evaluate Eq.~\eqref{eq_ahe_conduct}.

This interpolation approach suffers from a
band truncation error, because only Wannier interpolated
states are used to evaluate Eq.~\eqref{eq_ahe_conduct}.
However, the band truncation error has been shown to be
small in the case of AHE~\cite{PhysRevB.74.195118}
and also in the case of SHE~\cite{PhysRevX.11.041053}.

\subsection{Wannier interpolation of the spin and torque operators}
\label{eq_interpolate_spin_torque}
Within MFbSDFT, the
matrix elements of the spin operator
are defined by
\bege
\begin{aligned}
  \vn{S}_{\vn{k}nm}&=\frac{\hbar}{2}
  (\vn{\psi}^{\dagger}_{\vn{k}n\rightarrow},\vn{\psi}^{\dagger}_{\vn{k}n\leftarrow})
\begin{pmatrix}
  \vn{\sigma}&\vn{0}\\
  \vn{0} &\vn{0}
\end{pmatrix}
\begin{pmatrix}
  \vn{\psi}_{\vn{k}m\rightarrow}\\
  \vn{\psi}_{\vn{k}m\leftarrow}
\end{pmatrix}\\
&=\frac{\hbar}{2}
\vn{\psi}^{\dagger}_{\vn{k}n\rightarrow}
\vn{\sigma}
\vn{\psi}_{\vn{k}m\rightarrow}\\
&=\frac{\hbar}{2}
  (\vn{\psi}^{\dagger}_{\vn{k}n\rightarrow\uparrow},\vn{\psi}^{\dagger}_{\vn{k}n\rightarrow\downarrow})
  \vn{\sigma}
\begin{pmatrix}
  \vn{\psi}_{\vn{k}m\rightarrow\uparrow}\\
  \vn{\psi}_{\vn{k}m\rightarrow\downarrow}
\end{pmatrix}.
\end{aligned}  
\ee
In order to compute for example spin photocurrents~\cite{PhysRevResearch.4.043046}
from Wannier interpolation within MFbSDFT,
these matrix elements need to be interpolated.
We obtain the
interpolated $\tilde{\vn{S}}_{\vn{k}nm}$
similarly to Eq.~\eqref{eq_srnm} and Eq.~\eqref{eq_sknm}
(replace $\vn{s}_{\vn{k}nm}\rightarrow \vn{S}_{\vn{k}nm}$,
$\vn{s}_{\vn{R}nm}\rightarrow \vn{S}_{\vn{R}nm}$,
and
$\tilde{\vn{s}}_{\vn{k}nm}\rightarrow \tilde{\vn{S}}_{\vn{k}nm}$
in these equations).
Finally, 
the
interpolated $\tilde{\vn{S}}_{\vn{k}nm}$
may be transformed into the eigenbasis
similarly to Eq.~\eqref{eq_velo_inter_final}:
\bege\label{eq_spin_inter_final}
\langle \psi_{\vn{k}n\rightarrow} |  \vn{S} |\psi_{\vn{k}m\rightarrow}  \rangle=
\sum_{n'm'}\tilde{\vn{S}}_{\vn{k}n'm'}\mathcal{X}_{\vn{k}m'm}
\left[
\mathcal{X}_{\vn{k}n'n}
\right]^{*}.
\ee

The torque
operator $\vn{\mathcal{T}}$
is needed for the calculation
of the SOT~\cite{ibcsoit,invsot}.
At first glance, it is tempting to 
define the torque operator by
\bege\label{eq_ex_torque}
\vn{\mathcal{T}}_{\vn{k}nm}=
-\mu_{\rm B}\int d^{3}r
[\psi_{\vn{k}n\rightarrow}(\vn{r})]^{\dagger}\vht{\sigma}
\psi_{\vn{k}m\rightarrow}(\vn{r})\times \vn{\Omega}^{\rm xc}(\vn{r})
\ee
within MFbSDFT, where $\mu_{\rm B}$ is the Bohr magneton,
and $\vn{\Omega}^{\rm xc}(\vn{r})$
is the exchange field. However, the moment
potentials $\mathcal{V}^{(2+)}_{\sigma}(\vn{r})$
and $\mathcal{V}^{(3+)}_{\sigma}(\vn{r})$
(see Eq.~\eqref{eq_def_mi+})
may be spin-polarized in general, similarly to the
exchange  potential in the first moment. 
There is no convincing argument that 
one may substitute the exchange potential of the first
moment for $\vn{\Omega}^{\rm xc}(\vn{r})$ in Eq.~\eqref{eq_ex_torque}.
Instead, we expect that a suitable expression
for $\vn{\Omega}^{\rm xc}(\vn{r})$ may be derived
within the MFbSDFT framework, and that it will depend on the
potentials of the first, second, and third moments.

Therefore, we consider the alternative expression for
the torque operator
\bege\label{eq_soi_torque}
\vn{\mathcal{T}}_{\vn{k}nm}=
-\frac{i}{2}
\int d^{3}r
[\psi_{\vn{k}n\rightarrow}(\vn{r})]^{\dagger}
[H^{\rm SOI}(\vn{r}),\vn{\sigma}]
\psi_{\vn{k}m\rightarrow}(\vn{r}),
\ee
where $H^{\rm SOI}(\vn{r})$
is the SOI.
The torque operator may be interpolated analogously 
to the interpolation of the spin operator discussed above.

The torque operator may also be used to
compute the MAE~\cite{mae_torque_method,ibcsoit}.
Within MFbSDFT, the torque due to the magnetic
anisotropy is given by
\bege\label{eq_torque_for_mae}
\vn{T}^{\rm mae}
=-\frac{1}{\mathcal{N}}\sum_{\vn{k}n}
f_{\vn{k}n}
\langle
\psi_{\vn{k}n\rightarrow}
|
\vht{\mathcal{T}}
|
\psi_{\vn{k}n\rightarrow}
\rangle.
\ee

\section{Extension to more moments}
\label{sec_more}
In Ref.~\cite{momdis}
we present an efficient algorithm to construct
the spectral function from the first $2P$ spectral moment matrices,
where $P=1,2,3,\dots$.
The algorithm described in Ref.~\cite{momentis}, which we revisit
briefly in Sec.~\ref{sec_revisit_specfun}, is the special case with
$P=2$ of this more general algorithm.
We expect that the accuracy of the MFbSDFT approach can be enhanced
by increasing $P$. For example, in Ref.~\cite{momdis} we explain that it easy to
reproduce the jump of the momentum distribution function $n_{\vn{k}}$ of the
UEG
at $k_{\rm F}$ when $P\geq 3$, while this is difficult to achieve with $P=2$.

In Sec.~\ref{sec_mlwfs_from_moments} we describe the generation
of MLSMWFs when $P=2$. The extension to $P>2$ is straightforward.
As an example, consider the case $P=3$, i.e., assume that we construct
the spectral function from the first 6 spectral moment matrices
using the algorithm described in Ref.~\cite{momdis}.
In this case the poles of the spectral function
are the eigenvalues of a $3N_{\rm S}\times 3N_{\rm S}$
matrix $\vn{\mathcal{H}}_{\vn{k}}$. The eigenvectors
of $\vn{\mathcal{H}}_{\vn{k}}$ have $3N_{\rm S}$ components in this case
and they may be written in the form
\bege\label{eq_state_vectors_3mom}
\vn{\Psi}_{\vn{k}n}=\begin{pmatrix}
\vn{\psi}_{\vn{k}n\rightarrow}\\
\vn{\psi}_{\vn{k}n\nwarrow}\\
\vn{\psi}_{\vn{k}n\swarrow}
\end{pmatrix},
\ee
where $\vn{\psi}_{\vn{k}n\rightarrow}$, $\vn{\psi}_{\vn{k}n\nwarrow}$,
and $\vn{\psi}_{\vn{k}n\swarrow}$ are $N_{\rm S}$-component vectors.
$\vn{\psi}_{\vn{k}n\rightarrow}$ is the physical component from which the
charge density, the DOS, the spectral weights, and the expectation values of operators
can be computed. $\vn{\psi}_{\vn{k}n\nwarrow}$,
and $\vn{\psi}_{\vn{k}n\swarrow}$ are auxiliary components, which may be discarded
in a standard MFbSDFT selfconsistency loop after diagonalizing $\vn{\mathcal{H}}_{\vn{k}}$.
However, like in Sec.~\ref{sec_mlwfs_from_moments},
these auxiliary components need to be included into the generation of the
MLSMWFs.
Therefore, we construct the MLSMWFs from
\bege\label{eq_def_mlsmwfs_p3}
\begin{pmatrix}
\langle \vn{r}|\mathcal{W}_{\vn{R}n\rightarrow}\rangle\\
\langle \vn{r}|\mathcal{W}_{\vn{R}n\nwarrow}\rangle\\
\langle \vn{r}|\mathcal{W}_{\vn{R}n\swarrow}\rangle
\end{pmatrix}
=\frac{1}{\mathcal{N}}
\sum_{\vn{k}}\sum_{m=1}^{3N_{\rm S}} U_{mn}^{(\vn{k})} e^{-i\vn{k}\cdot \vn{R}}
\begin{pmatrix}
  \psi_{\vn{k}m\rightarrow}(\vn{r})\\
  \psi_{\vn{k}m\nwarrow}(\vn{r})\\
   \psi_{\vn{k}m\swarrow}(\vn{r})
\end{pmatrix}
\ee
in this case. Here, $\vn{U}^{(\vn{k})}$
is a $3N_{\rm S}\times N_{\rm W}$ matrix.

\section{Applications}
\label{sec_results}

When the magnetization is along the
[001] direction, GGA predicts the intrinsic
AHE in Ni to be -2200~S/cm,
which is significantly larger than
the experimental value of -646~S/cm~\cite{PhysRevB.84.144427}.
Using GGA+$U$ with
$U=1.9$~eV, one obtains the intrinsic 
AHE of -1066~S/cm~\cite{PhysRevB.84.144427}.
The remaining discrepancy between experiment and theory
is 420~S/cm. This discrepancy can be explained by
the side-jump AHE~\cite{PhysRevLett.107.106601}.

MFbSDFT may be used to reproduce the experimental values of
the exchange splitting, of the band width, and of the valence
band satellite position in fcc Ni~\cite{momentis,momdis}.
To compute the AHE in Ni from MLSMWFs,
we first perform self-consistent MFbSDFT calculations with SOI.
We perform these calculations with various different  $d^{(2+)}$
parameters in the range 15-20 to investigate the dependence of the AHE on   $d^{(2+)}$,
i.e.\ we use Eq.~\eqref{eq_expand_vi+_vc}, but we set $d^{(3+)}=0$.
In order to keep the magnetic moment fixed at around 0.6~$\mu_{\rm B}$, which
is the value measured in experiments, we need to
spin-polarize $\mathcal{V}^{(2+)}$. We
use $\mathcal{V}^{(2+)}_{\sigma}=\zeta^t_{\sigma}\mathcal{V}^{(2+)}$,
where $\zeta_{\sigma}=(1-\sigma (n_{\uparrow}-n_{\downarrow})/n)$, and $t$ is
determined at every value of $d^{(2+)}$ to match the experimental magnetic moment.
Next, we compute the
matrix elements $M_{mn}^{(\vn{k},\vn{b})}$
and $A_{mn}^{(\vn{k})}$ as discussed in Sec.~\ref{sec_overlaps}, Sec.~\ref{sec_initial_proj},
and Sec.~\ref{sec_mlsmwfs_soi}.
We generate MLSMWFs using the {\tt wannier90} code~\cite{wannier90communitycode}
and disentanglement,
where
we set the lower bound of the frozen window at around 80~eV below the Fermi energy 
and the upper bound at around 4~eV above the Fermi energy.
We construct 36 spinor MLSMWFs 
from 72 MFbSDFT bands.

\begin{figure}
\includegraphics[angle=0,width=\linewidth]{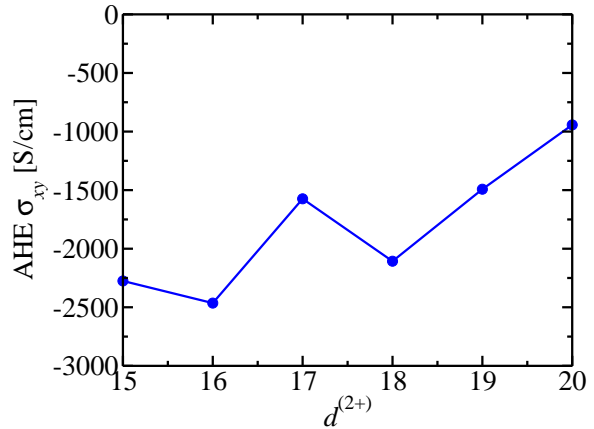}
\caption{\label{fig_AHE_Ni}
  AHE conductivity $\sigma_{xy}$ vs.\
  the prefactor $d^{(2+)}$
of the second moment potential.
}
\end{figure}

In Fig.~\ref{fig_AHE_Ni}
we plot the AHE
obtained from MLSMWFs
as explained in Sec.~\ref{sec_interpolate_mlsmwfs_ahe}
as a function
of the prefactor $d^{(2+)}$ used
in the potential
of the second moment.
With increasing $d^{(2+)}$ the magnitude of
$\sigma_{xy}$ decreases. At $d^{(2+)}=20.0$
the intrinsic AHE is -1000~S/cm.
If we assume that the side-jump contribution
to the AHE is around 
 400~S/cm~\cite{PhysRevLett.107.106601}, this is
in good agreement with the experimental
value of -646~S/cm.

\begin{figure}
\includegraphics[angle=0,width=\linewidth]{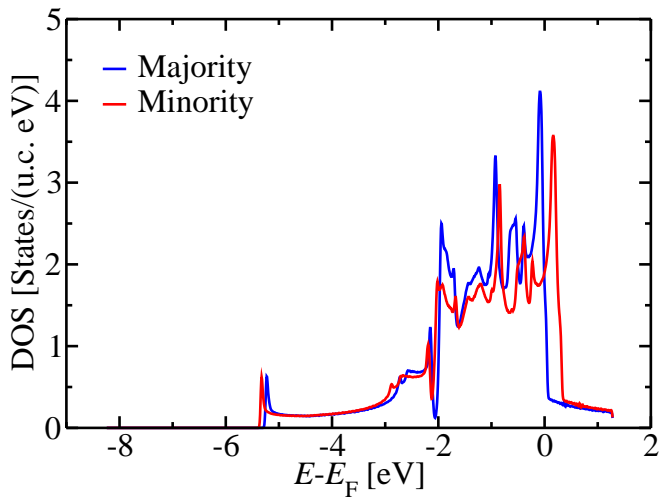}
\caption{\label{fig_Ni_DOS}
  DOS in fcc Ni obtained from MFbSDFT.
}
\end{figure}

In Ref.~\cite{momentis} we used $d^{(2+)}=15.0$
in order to reproduce the experimental bandwidth,
exchange splitting, and position of the satellite peak.
However, using $d^{(2+)}=20.0$ instead reproduces these
experimental features also quite well, which we show in
Fig.~\ref{fig_Ni_DOS}.
In Ref.~\cite{momdis} we have found that the valence band
satellite is in much better agreement with DMFT calculations and with experiment
if the third moment potential is computed from the second moment
potential using the constraint of the momentum distribution function
of the UEG. However, since we have currently developed this procedure
only for the UEG without spin-polarization we needed to apply
a similar spin-polarization factor $\zeta^t_{\sigma}$ like in the present
calculations. As a result, the spectral density of Ni in Ref.~\cite{momdis}
matches experiment concerning the spin-polarization of the satellite peak,
and the band width of the main band. However, it suffers from a similar
overestimation of the exchange splitting as standard KS-DFT with LDA.
In contrast, the present calculation yields the exchange-splitting close to experiments.
Since the AHE depends strongly on the Fermi surface~\cite{PhysRevLett.93.206602,PhysRevB.76.195109,PhysRevB.89.117101}
we therefore use here the simpler approach of Eq.~\eqref{eq_expand_vi+_vc}
instead of the improved approach of Ref.~\cite{momdis}.

\section{Summary}
\label{sec_summary}
We describe the construction of Wannier functions from
the first 4 
spectral moment matrices. We show that these MLSMWFs
can be used for the efficient interpolation of material
property tensors such as the AHE within MFbSDFT.
 This paves the way for the application
of MFbSDFT to compute response properties of materials.
We demonstrate that MFbSDFT is able to reproduce the
experimentally measured AHE in fcc Ni,
similarly to LDA+$U$.
Finally, we discuss that MLSMWFs may be computed also from
the first 6 moments, and generally from the first $2P$ moments. 
This opens the perspective of using as many moments as necessary
to reproduce all spectral features accurately in MFbSDFT.

\section*{Acknowledgments}
The project is funded by the Deutsche
Forschungsgemeinschaft (DFG, German Research Foundation) $-$ TRR 288 $-$ 422213477 (project B06),
CRC 1238, Control and Dynamics of Quantum Materials: Spin
orbit coupling, correlations, and topology (Project No. C01),
SPP 2137 ``Skyrmionics",  and Sino-German research project
DISTOMAT (DFG project MO \mbox{1731/10-1}).
We also
acknowledge financial support from the European Research
Council (ERC) under the European Union’s Horizon 2020
research and innovation program (Grant No. 856538, project
``3D MAGiC'') and computing resources granted by the Jülich Supercomputing
Centre under
project No. jiff40.

\bibliography{momwan}

\end{document}